\pdfoutput=1
\documentclass[11pt]{article}

\usepackage{xcolor}
\usepackage{booktabs}

\usepackage{acl}
\usepackage{times}
\usepackage{latexsym}

\usepackage{graphicx}
\usepackage{caption}
\usepackage{changepage}
\usepackage{booktabs}
\usepackage{subcaption}
\usepackage{algorithm}
\usepackage{algpseudocode}
\usepackage{mdframed}

\usepackage[T1]{fontenc}

\usepackage[utf8]{inputenc}
\usepackage{acl}
\usepackage{makecell}

\AtBeginDocument{%
 }

\usepackage{subcaption}
\usepackage{float}
\usepackage{microtype}

\usepackage{inconsolata}
\begin{document}

\title{PseudoSeer: a Search Engine for Pseudocode}


\author{%
  Levent Toksoz$^{1}$ \quad Mukund Srinath$^{2}$ \quad Gang Tan$^{1}$ \quad C. Lee Giles$^{1,2}$ \\
  \\
  \begin{tabular}{c@{\hskip 1.5cm}c}
    $^{1}$\makecell{The Pennsylvania State University, \\ Computer Science and Engineering, \\ University Park, Pennsylvania, 16802} & 
    $^{2}$\makecell{The Pennsylvania State University, \\ Information Sciences and Technology, \\ University Park, Pennsylvania, 16802} \\
  \end{tabular} \\
  \texttt{lkt5297@psu.edu} \quad \texttt{mus824@psu.edu} \quad \texttt{gtan@psu.edu} \quad \texttt{clg20@psu.edu}
}

\maketitle

\begin{abstract}
A novel pseudocode search engine is designed to facilitate efficient retrieval and search of academic papers containing pseudocode. By leveraging Elasticsearch, the system enables users to search across various facets of a paper, such as the title, abstract, author information, and LaTeX code snippets, while supporting advanced features like combined facet searches and exact-match queries for more targeted results. A description of the data acquisition process is provided, with arXiv as the primary data source, along with methods for data extraction and text-based indexing, highlighting how different data elements are stored and optimized for search. A weighted BM25-based ranking algorithm is used by the search engine, and factors considered when prioritizing search results for both single and combined facet searches are described. We explain how each facet is weighted in a combined search. Several search engine results pages are displayed. Finally, there is a brief overview of future work and potential evaluation methodology for assessing the effectiveness and performance of the search engine is described.
\end{abstract}

\section{Introduction}

The ever-growing volume of academic literature necessitates efficient search tools that cater to the specific needs of researchers. Within this domain, code plays a crucial role in various disciplines, serving as a foundation for replicating experiments, validating methodologies, and promoting understanding. Traditional academic search engines, however, often lack the functionality to effectively search for code within research papers. In this paper, we present PseudoSeer, a pseudocode search engine capable of searching over pseudocode in academic documents. 

PseudoSeer bridges the gap between traditional text-based search and the need to locate code snippets embedded within academic publications. Pseudocode uses a language that closely resembles programming languages but focuses on readability and general concepts over syntax specifics, allowing researchers to represent algorithms and data structures in an easy-to-understand manner. A dedicated search engine capable of indexing and retrieving pseudocode offers a significant advantage by enabling targeted searches based on the inherent logic and functionality described within a paper.

This paper details the design and implementation of a pseudocode search engine built using the robust capabilities of Elasticsearch. The system empowers users to search across various facets of a research paper, including the title, abstract, author information, and LaTeX code sections, as well as perform combined searches across multiple facets. This functionality caters to diverse search scenarios. For instance, a researcher might utilize the engine to locate papers implementing a specific algorithm or explore the use of a particular data structure across various publications. 

The development of this system involved several key considerations. Dataset is carefully chosen, with Arxiv serving as the primary source for academic papers containing code, ensuring both extensive coverage and diversity to handle broad range of  search queries. A critical component we discuss in the paper is the indexing strategy, where decisions regarding how different data elements (text, code snippets, author information) are stored and optimized for search significantly impact retrieval effectiveness. Another core aspect is the ranking algorithm, which determines the order in which search results are presented to the user. Factors such as performing single or combined facet searches, along with the relevance of each facet, all play a role in prioritizing the results.

Developing a pseudocode search engine presents both opportunities and challenges. On one hand, the system offers a powerful tool for code-centric literature exploration. However, complexities arise in accurately identifying and parsing code sections within academic papers. Additionally, tradeoffs exist between comprehensiveness and search performance. Indexing a vast amount of code data can be computationally expensive, while focusing solely on keywords within code snippets might miss the underlying logic and functionality. This paper addresses these challenges by outlining the chosen approaches and discussing the inherent difficulties involved. In summary, our contributions are: 
\begin{itemize}
    \item \textbf{Development of PseudoSeer:} A specialized search engine for pseudocode retrieval
    \item \textbf{Facet-Based Search Functionality:} Support for combined facet searches across title, abstract, references, author, and pseudocode for refined queries.
    \item \textbf{Exact-Match Search:} Exact-match query functionality using quotation marks.
    \item \textbf{Weighted Ranking Mechanism:} Implementation of a BM25-based ranking system with facet-specific weighting to enhance relevance in search results.
\end{itemize}

\section{Related Work}
Several pseudocode datasets have been developed to support research in various code related translation and generation tasks. Notably, the SPOC dataset \citet{Spoc} contains around 20,000 programs with human-authored pseudocodes and test cases, designed for translating pseudocode to C++ code. Another significant dataset is by \citet{Oda2015LearningTG}, which includes approximately 16,000 manually crafted pseudocodes. The dataset is used for statistical machine translation. Additionally, \citet{zavershynskyi2018naps} provide a dataset of about 2,000 manually written pseudocodes for program synthesis tasks.

While these datasets are valuable for specific tasks, they represent a limited variety of pseudocodes and are hand-crafted by programmers. Moreover, their relatively small size and lack of diversity make them less suitable for use in search engines, which require more extensive and varied datasets to effectively handle a broad range of pseudocode search queries. To that end, \citet{Toksoz2024} provide a much larger dataset containing 320,000 pseudocode samples extracted from scholarly papers on arXiv, which will serve as the foundation for the dataset we use to build our pseudocode search engine.

Several search engines have tackled similar challenges in retrieving structured content, particularly in domains that involve specialized formats similar to pseudocode. For instance, \citet{kohlhase2006search} and \citet{kohlhase2012mathwebsearch} developed a math formula search engine called MathWebSearch, which structurally indexes mathematical expressions to enhance retrieval efficiency. Further, \citet{normann2007extended} introduced extended normalization techniques aimed at mathematical formulas to detect not only structurally identical queries but also logically equivalent ones, regardless of the nomenclature chosen.

In addition to search engines, various efforts have focused on extracting structured information from documents, employing methods designed to retrieve elements such as mathematical equations, tables, figures, and metadata. These works demonstrate the potential of machine learning models to accurately extract structured content from PDFs, yet there remains a gap in addressing the unique challenges of generating pseudocode datasets from scholarly sources like arXiv. Recent advancements in this area include methods developed by \citet{DBLP:journals/corr/abs-2004-14356}, who built a machine learning pipeline for extracting results from research paper, and \citet{nassar2022tableformer}, who introduced TableFormer, a model specifically designed to capture table data. Similarly, \citet{9085944} present techniques for chart extraction, aiming to improve the accessibility of visual data in academic texts. Other notable contributions, such as  and \citet{10.1145/1065385.1065418}, focus on math equation detection, while \citet{DBLP:journals/corr/abs-2003-08005}, \citet{houyufang2019acl} and \citet{houyufang2021eacl} address the extraction of tasks, datasets, and evaluation metrics as distinct entities. \citet{blecher2023nougat} and tools like \citet{GROBID} further demonstrate advancements in general metadata extraction and 
scientific document processing. 

\section{Data Collection}

The pseudocode search engine utilizes data provided by Toksoz et al. (2024)\footnote{The dataset can be accessed via the \href{https://github.com/letoksoz/arxiv-pseudocode/}{\textit{arxiv-pseudocode}} repository on GitHub.}, which is extracted from the LaTeX files of scholarly papers on arXiv, starting with year 1991 and ending in June of 2023. These files are stored across both Amazon S3 buckets and Google Cloud. To gather the LaTeX files, their extraction process scanned over 10 TB of arXiv data, recursively extracting ZIP files and matching the extracted content to the corresponding papers using arXiv identifiers. They also stored all the metadata information such as the arXiv identifier, any equations referenced by the pseudocode, and the year it was stored in arXiv. The pseudocode extraction process operates as follows: first, it identifies the \texttt{\textbackslash begin\{algorithm\}} and \texttt{\textbackslash end\{algorithm\}} tags within the LaTeX files of scholarly papers on arXiv. Next, it extracts the sections enclosed by these tags, recognizing them as pseudocode. In total nearly $320,000$ pseudocode examples over $2.2$ million scholarly papers are extracted. 

The pseudocode dataset obtained from Toksoz et al. (2024) is further processed to include additional metadata information such as arXiv weblinks of the papers. Following the Reference Detection Algorithm described by Toksoz et al., references are extracted by matching labels with their associated 'ref' tags using regular expressions, and capturing a window of text around the referenced section. The extracted reference snippets are then further processed by shortening and cleaning the LaTeX syntax as much as possible to provide a more accessible overview and improved search experience. It should be noted that not all pseudocode references could be extracted, particularly those lacking associated reference tags or having tags that are too complex for regular expressions to handle. Additionally, papers and pseudocode that could not be parsed and indexed without errors were omitted, such as those with incomplete LaTeX code.

\section{Search Interface}
\begin{figure}[H]
    \centering
    \includegraphics[width=0.7\textwidth]{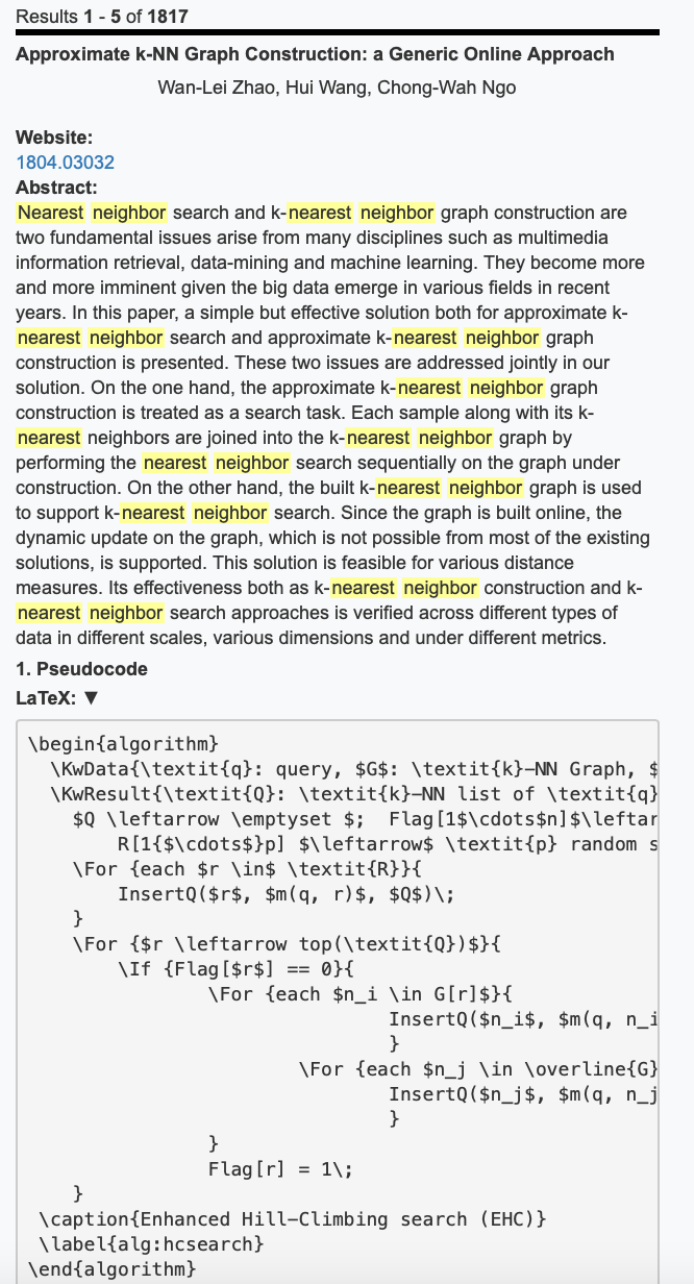}
    \caption{Results page obtained by using the search query 'nearest neighbor' in the abstract field}
    \label{fig:results}
\end{figure}

\begin{figure}[h]
    \centering
    \includegraphics[width=0.7\textwidth]{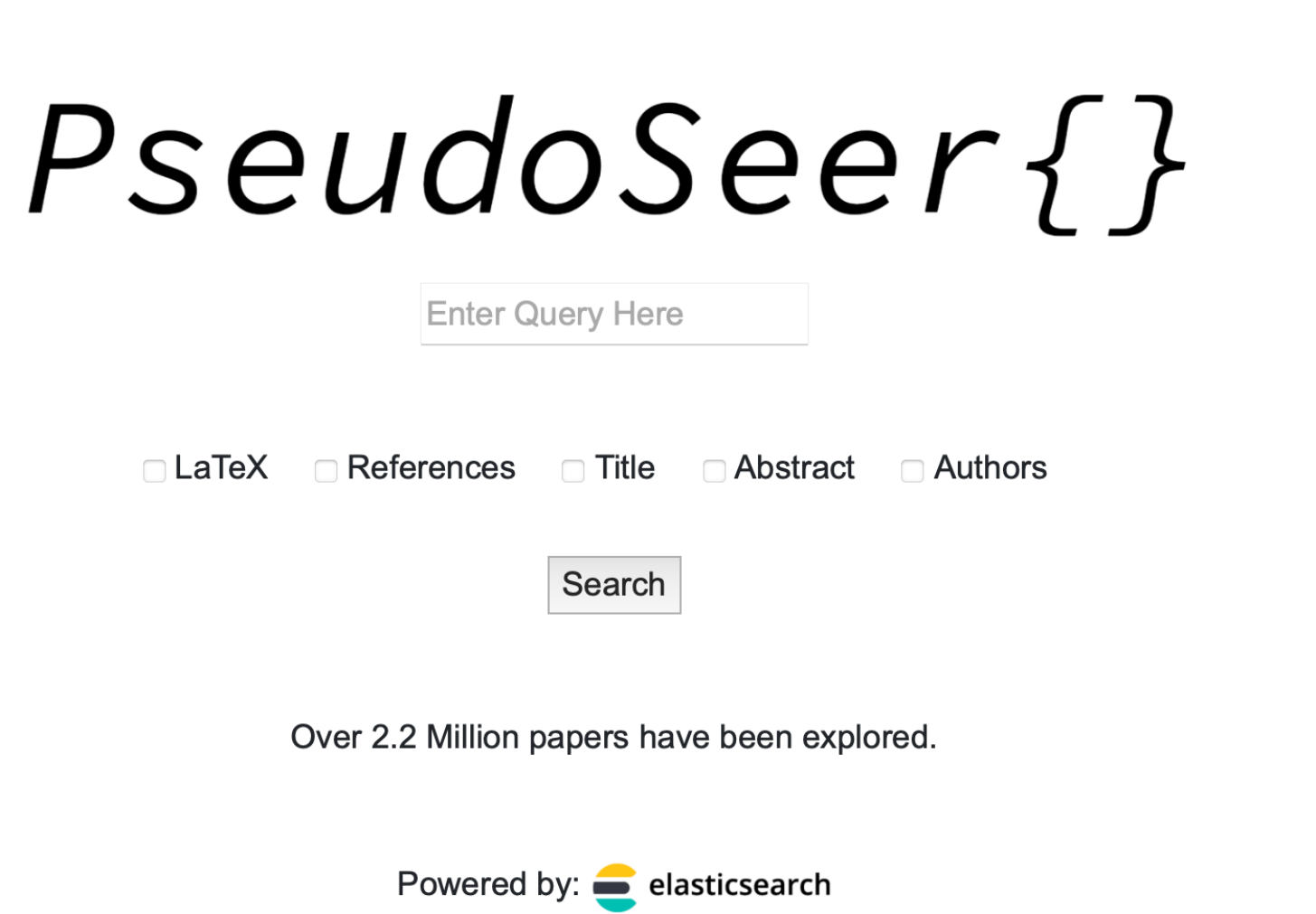}
    \caption{Landing page with options to search in LaTeX, references, title, abstract, and authors, individually or combined.}
    \label{fig:landing}
\end{figure}

The pseudocode search engine\footnote{The search engine can be accessed via the \href{https://pseudoseer.ist.psu.edu/}{\textit{pseudoseer}} link.} prioritizes a user-friendly interface that resembles common search engines to ensure ease of use. The landing page presents a search bar where users can enter their queries. Figures \ref{fig:landing} and \ref{fig:results} illustrate these features. Unlike traditional search engines, however, this interface offers additional functionalities to refine the search based on the specific elements within a research paper. The page provides a set of radio buttons that allow users to specify which part of the paper they want to search, with options including title, abstract, author, or LaTeX code sections. Users can also select multiple buttons simultaneously to perform a combined search. By default, if no buttons are selected, a combined search across all fields is conducted. The interface also supports exact query matching by allowing users to wrap quotation marks around the desired query. This functionality caters to researchers with diverse search goals. For instance, a user aiming to locate papers implementing a specific algorithm might choose to search within the code sections, while someone interested in an author's code contributions could utilize the author search option.

The search results page displays a list of retrieved papers, ordered by a custom ranking function discussed in the following section. Each entry on the results page provides relevant information about the retrieved paper, including the title, arXiv website, authors, and a snippet of text containing the matched keywords. The system also highlights the specific sections of the paper where the search terms were found (title, abstract, code snippet). This visual cue can be beneficial for users by allowing them to quickly assess if the retrieved paper aligns with their search intent. Furthermore, the interface offers options for pagination, enabling users to explore a larger set of results if the initial search retrieves a significant number of papers. 

\section{Tokenization and Indexing}
The effectiveness of the pseudocode search engine hinges on a robust indexing strategy that efficiently stores and retrieves information from various elements within a research paper. This section delves into the process of indexing different data types, with a particular focus on the complexities involved in handling both LaTeX pseudocode and its cleaned references.

Standard text processing techniques are employed for indexing the title, abstract, and author information of research papers. This involves tokenization, where the text is broken down into individual words or meaningful units. We tokenized each of the above mentioned fields using grammar based tokenization that works based on the Unicode text segmentation algorithm.

\subsection{Indexing Code}

Indexing code sections, especially those written in LaTeX, presents a unique challenge. LaTeX, a document typesetting system, utilizes a combination of plain text and markup commands to define the structure and formatting of the document. A straightforward approach involves converting the entire LaTeX code to plain text before performing tokenization. This simplifies the indexing process but can lead to loss of code-specific constructs that are essential for meaningful search results. For instance, when commands such as "\texttt{\textbackslash for}" or "\texttt{\textbackslash begin\{algorithm\}}" are removed, structure and logical flow in the pseudocode can be compromised. 

In our search engine, we retain the LaTeX code itself for indexing, treating it as regular text while keeping important commands intact to preserve the structural cues within the pseudocode. This allows us to capture specific elements like "\texttt{\textbackslash for}" loops and "\texttt{\textbackslash If}" conditions directly, without breaking down the LaTeX into plain text and risking a loss of context and structure, especially since there are a considerable number of LaTeX pseudocode examples that are like  Figure~\ref{example}. 

\begin{figure}[H]
\centering
\begin{mdframed}[roundcorner=10pt, linewidth=1pt]
\begin{algorithmic}
\State \texttt{\textbackslash For} each element in the list
    \State \hspace*{1.5em} \texttt{\textbackslash If} element satisfies condition
        \State \hspace*{3em} perform action
    \State \hspace*{1.5em} \texttt{\textbackslash EndIf}
\State \texttt{\textbackslash EndFor}
\end{algorithmic}
\end{mdframed}
\caption{Example pseudocode with for and if statements.}
\label{example}
\end{figure}

We also index the references surrounding each pseudocode snippet to enable keyword-based searches within the pseudocode’s descriptive context, while avoiding complex LaTeX syntax. This dual-layered indexing is particularly useful for users who may not know or prefer not to use the exact code syntax, as it allows them to locate pseudocode based on broader thematic keywords and descriptions found in the surrounding text, rather than being restricted to the code-construct-heavy language within the pseudocode itself. While this strategy requires more storage and adds complexity to the retrieval and ranking process, it enhances the search engine's flexibility and precision for both structural and contextual searches.



\section{Ranking}
Another crucial factor that affect the effectiveness of the pseudocode search engine is an accurate and efficient ranking mechanism that prioritizes the most relevant results for a given query. This section details the ranking process employed by the search engine. For single field searches, the search engine utilizes the BM25 algorithm, which is based on TF-IDF (Term Frequency-Inverse Document Frequency) metric. The BM25 algorithm considers the frequency of query terms in a document and adjusts the score based on the rarity of those terms across all indexed documents, while applying a saturation effect to prevent overly frequent terms from dominating the score. The score is also normalized by the document length to ensure that more relevant results are ranked higher while accounting for natural variations in term frequency and document length. 

In multi-field searches, the search engine assigns predetermined weights to each field. These weights are then incorporated into the BM25 algorithm to calculate the weighted scores. The weights are chosen based on the potential relevance of the fields in real-world use cases. For instance, results obtained by searching in the LaTeX pseudocode and authors fields are likely to be less relevant to the user than those from the abstract and title. To that end, the weights for the LaTeX pseudocode and authors fields are set to one, while the weights for other fields are set to two.

\section{Evaluation}
\begin{figure}[H]
    \centering
    \begin{subfigure}[b]{0.9\textwidth}
        \centering
        \includegraphics[width=\textwidth]{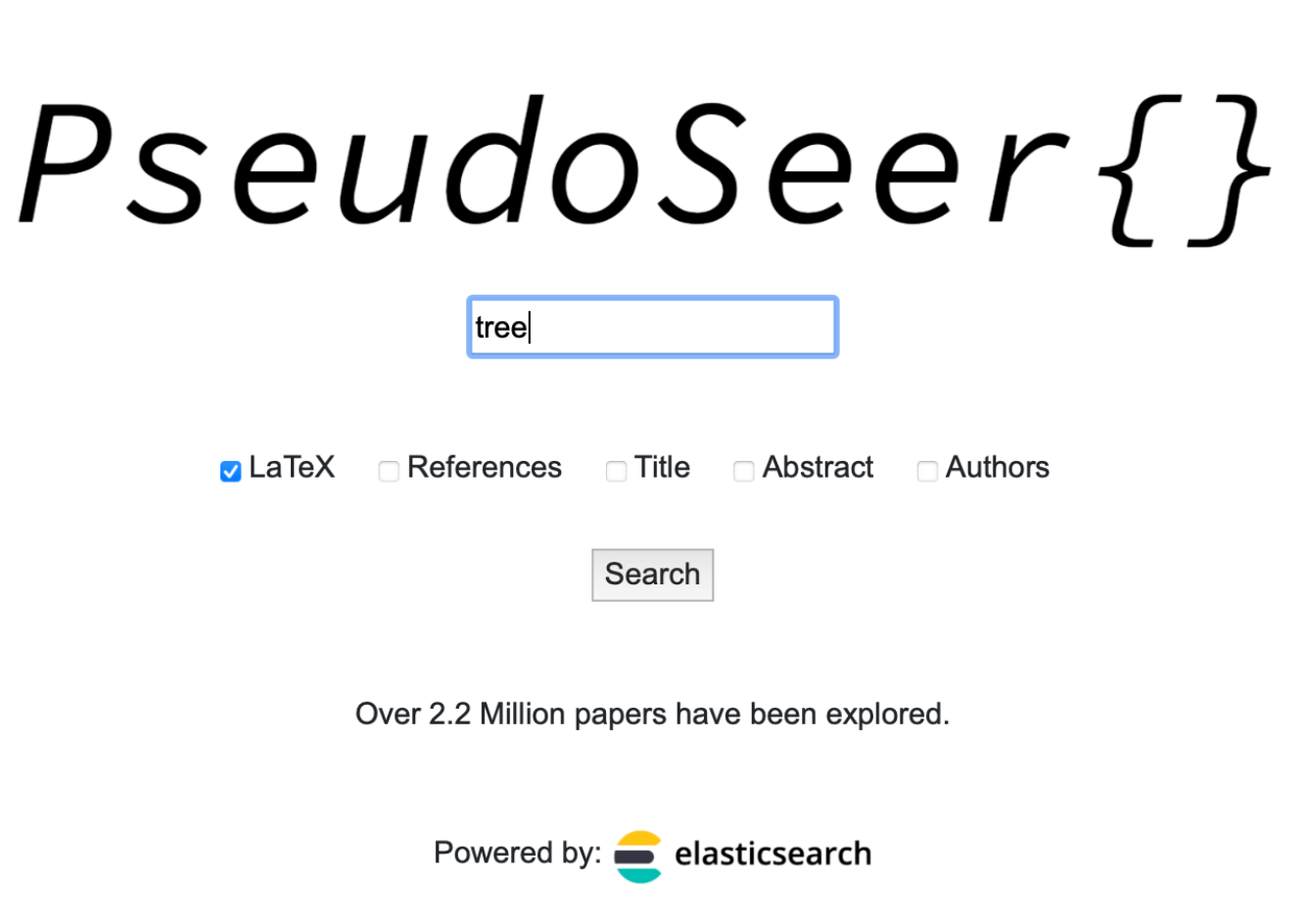}
        \caption{Single Field Search}
        \label{fig:singelanding}
    \end{subfigure}
    \hfill
    \begin{subfigure}[b]{0.7\textwidth}
        \centering
        \includegraphics[width=\textwidth]{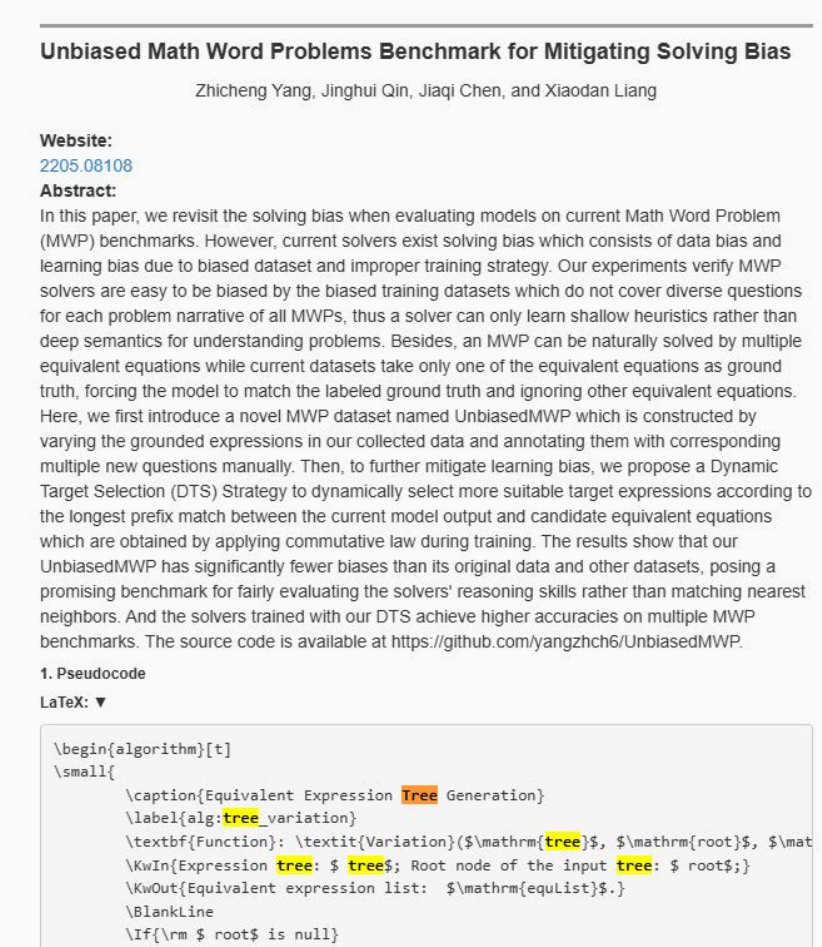}
        \caption{Results page for Single Field}
        \label{fig:singresults}
    \end{subfigure}
    \caption{Single-field search using a search query 'tree' and the search field LaTeX}
    \label{fig:single}
\end{figure}

\begin{figure}[H]
    \centering
    \begin{subfigure}[b]{0.9\textwidth}
        \centering
        \includegraphics[width=\textwidth]{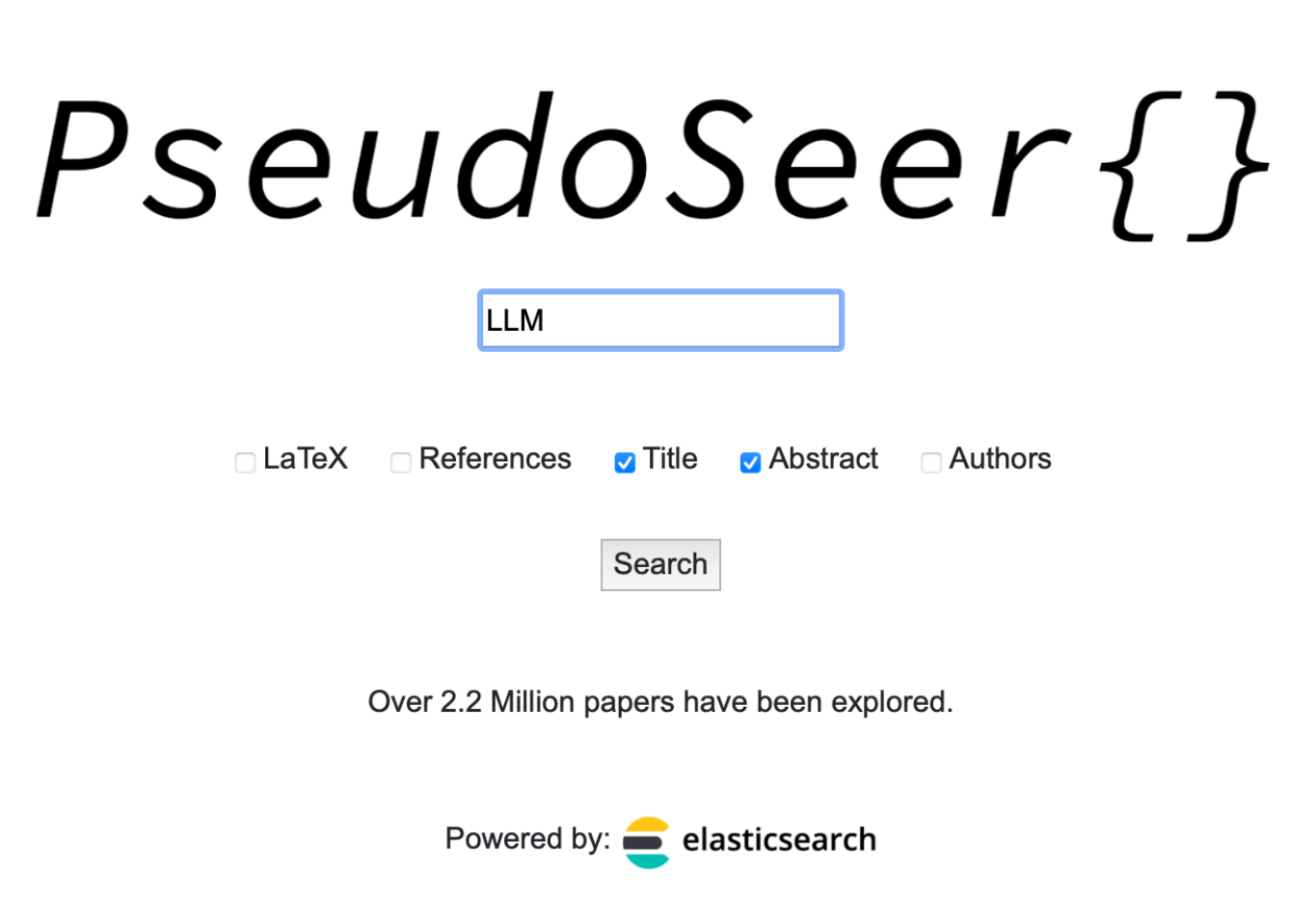}
        \caption{Combined search}
        \label{fig:multisearch}
    \end{subfigure}
    \hfill
    \begin{subfigure}[b]{0.7\textwidth}
        \centering
        \includegraphics[width=\textwidth]{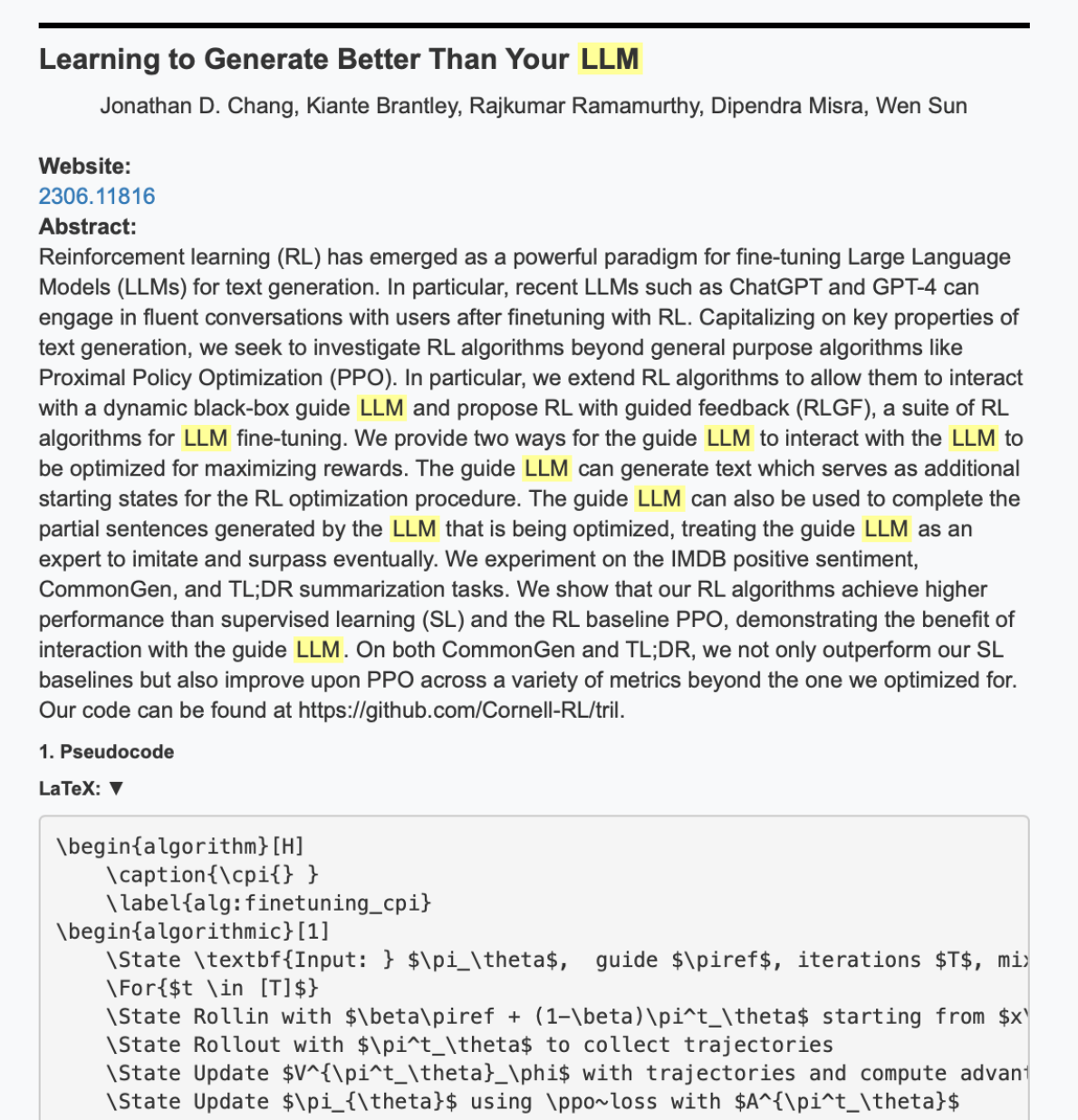}
        \caption{Results page for combined search}
        \label{fig:multilanding}
    \end{subfigure}
    \caption{Combined search using the search query 'LLM' across the title and abstract fields}
    \label{fig:combined_search}
\end{figure}

\begin{figure}[H]
    \centering
    \begin{subfigure}[b]{0.9\textwidth}
        \centering
        \includegraphics[width=\textwidth]{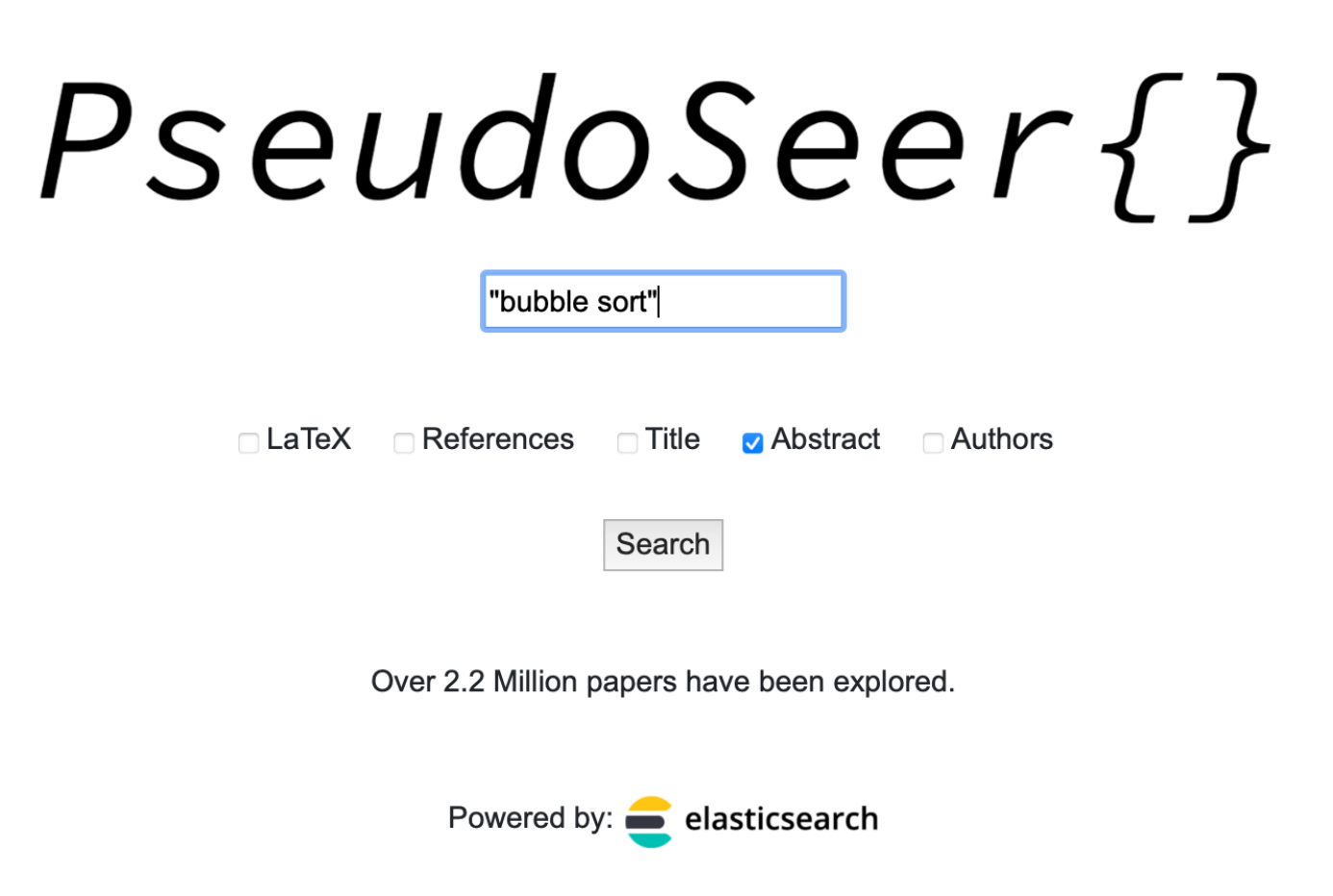}
        \caption{Exact Query Matching}
        \label{fig:exactlanding}
    \end{subfigure}
    \hfill
    \begin{subfigure}[b]{0.7\textwidth}
        \centering
        \includegraphics[width=\textwidth]{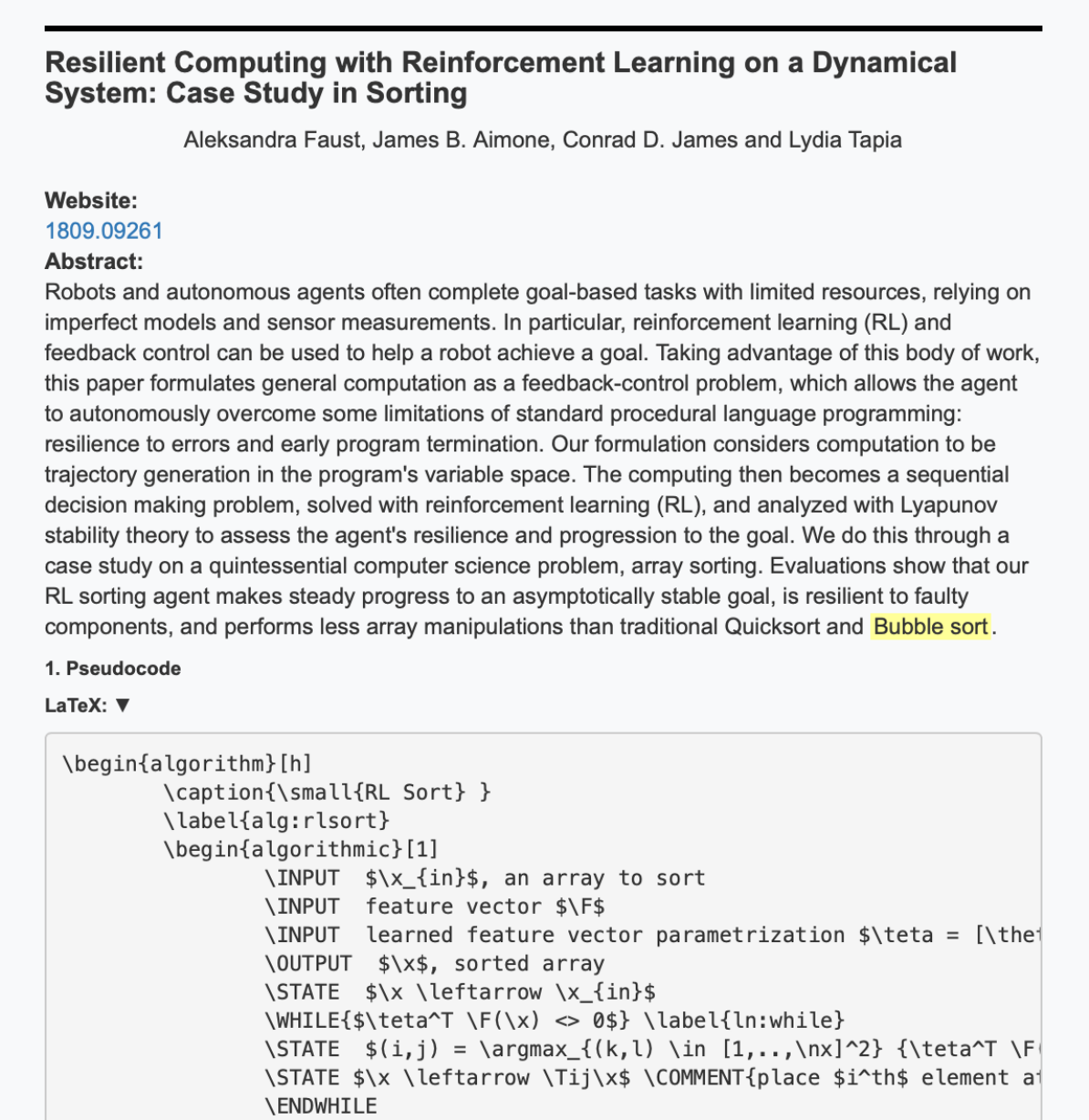}
        \caption{Results page for Exact Match}
        \label{fig:exactresults}
    \end{subfigure}
    \caption{Exact query matching with the phrase 'bubble sort' in quotation marks, searched within the abstract field}
    \label{fig:exact}
\end{figure}

Currently, the search engine evaluation is conducted by manually inspecting the results obtained from various query types. Figure \ref{fig:single} illustrates a single-field search and the corresponding results displayed by the search engine. Figure \ref{fig:combined_search} demonstrates a combined-field search, showing how multiple fields can be queried simultaneously with the results displayed accordingly. Similarly, Figure \ref{fig:exact} presents an exact-match search query and its respective results as produced by the search engine. Enhanced evaluation will be a focus of future.

\section{Future Work}
We briefly discuss potential improvements for the search engine. Currently, the data only consists of pseudocode from papers containing \texttt{\textbackslash begin\{algorithm\}} and \texttt{\textbackslash end\{algorithm\}} tags. A machine learning-based solution could be implemented to detect additional patterns and extract pseudocode, thereby expanding the dataset. Additionally, the search functionality could be enhanced by using a BERT-based model to better match user queries with the desired pseudocode. Moreover, incorporating different ranking algorithms that can dynamically adapt to both single-field and multi-field search queries could improve the precision of results based on the query context. Additional improvements could include integrating LaTeX pseudocode rendering into the interface, along with creating a more comprehensive evaluation framework to measure the effectiveness of ranking algorithms and user experience enhancements.


\section{Conclusion}
A novel pseudocode search engine is introduced and built with Elasticsearch. It enables efficient searching across various facets of research papers containing pseudocode. These facets include the title, abstract, author information, and LaTeX code sections, providing a comprehensive tool for locating relevant academic content. Users can choose to search within a single field or combine multiple facets for more refined results. Additionally, the ranking algorithm is designed to prioritize relevant search results based on whether the user is performing a single-field or multi-field search, while also considering the relevance of each field in multi-field searches.

\newpage

\section{Acknowledgements}
The arXiv is gratefully acknowledged for providing access to documents with pseudocode and their latex versions.





\end{document}